\documentclass{aastex631}

\accepted{October 18, 2024}

\begin{document}

\title{Database of Candidate Targets for the LIFE Mission}

\author[0009-0006-0336-6827]{Franziska Menti}
\affiliation{Institute for Particle Physics and Astrophysics, ETH Zurich, Wolfgang-Pauli-Str. 27, 8093 Zurich, Switzerland}

\author[0000-0002-7349-1387]{Jos\'e~A. Caballero}
\affiliation{Centro de Astrobiolog\'ia, CSIC-INTA, Camino Bajo del Castillo s/n, ESAC campus, 28692 Villanueva de la Ca\~nada, Madrid, Spain}

\author[0000-0001-9064-5598]{Mark~C. Wyatt}
\affiliation{Institute of Astronomy, University of Cambridge, Madingley Road, Cambridge CB3 0HA, UK}

\author[0000-0003-1756-4825]{Antonio Garc\'ia Mu\~noz}
\affiliation{Universit\'e Paris-Saclay, Universit\'e Paris Cit\'e, CEA, CNRS, AIM, 91191 Gif-sur-Yvette, France}

\author[0000-0002-3481-9052]{Keivan~G. Stassun}
\affiliation{Department of Physics and Astronomy, Vanderbilt University, Nashville, TN 37235, USA}

\author[0000-0002-0006-1175]{Eleonora Alei}
\affiliation{NPP Fellow, NASA Goddard Space Flight Center, 8800 Greenbelt Rd, 20771, Greenbelt, MD, USA}

\author{Markus Demleitner}
\affiliation{Zentrum f\"ur Astronomie der Universit\"at Heidelberg, Astronomisches Rechen-Institut, M\"onchhofstr. 12--14, 69120 Heidelberg, Germany}

\author[0000-0001-6831-7547]{Grant Kennedy}
\affiliation{Department of Physics, University of Warwick, Gibbet Hill Road, Coventry CV4 7AL, UK}

\author[0000-0002-3286-7683]{Tim Lichtenberg}
\affiliation{Kapteyn Astronomical Institute, University of Groningen, P.O. Box 800, 9700 AV Groningen, The Netherlands}

\author[0000-0002-4658-0616]{Uwe Schmitt}
\affiliation{Scientific IT Services, ETH Z\"urich, Binzm\"uhlestr. 130, 8092 Z\"urich, Switzerland}

\author[0000-0002-1043-8853]{Jessica S. Schonhut-Stasik}
\altaffiliation{Neurodiversity Inspired Science and Engineering Fellow}
\affiliation{Physics \& Astronomy Department,
Vanderbilt University,
6301 Stevenson Center,
Nashville, TN 37235}

\author[0000-0001-8618-3343]{Haiyang~S. Wang}
\affiliation{Center for Star and Planet Formation, Globe Institute, University of Copenhagen, $\O$ester Voldgade 5-7, 1350 Copenhagen, Denmark}

\author[0000-0003-3829-7412]{Sascha~P. Quanz}
\affiliation{Institute for Particle Physics and Astrophysics, ETH Zurich, Wolfgang-Pauli-Str. 27, 8093 Zurich, Switzerland}
\affiliation{Department of Earth and Planetary Sciences, ETH Zurich, Sonneggstrasse 5, 8092 Zurich, Switzerland}

\collaboration{20}{and the LIFE Collaboration}

\begin{abstract}

We present the database of potential targets for the Large Interferometer For Exoplanets (LIFE), a space-based mid-infrared nulling interferometer mission proposed for the Voyage 2050 science program of the European Space Agency (ESA). The database features stars, their planets and disks, main astrophysical parameters, and ancillary observations. It allows users to create target lists based on various criteria to predict, for instance,  exoplanet detection yields for the LIFE mission. 
As such, it enables mission design trade-offs, provides context for the analysis of data obtained by LIFE, and flags critical missing data. Work on the database is in progress, but given its relevance to LIFE and other space missions, including the Habitable Worlds Observatory (HWO), we present its main features here.  
A preliminary version of the LIFE database is publicly available on the German Astrophysical Virtual Observatory (GAVO). 
\end{abstract}

\section{Introduction} \label{sec:intro}

The LIFE mission \citep{2022A&A...664A..21Q} is designed to address one of the long-term goals of exoplanet science: the detailed  atmospheric characterization of dozens of temperate terrestrial exoplanets. In the current configuration, LIFE shall consist of four collector spacecraft, separated between tens to hundreds of meters, and a beam combined spacecraft. LIFE will search for atmospheric biosignatures and will revolutionize our general understanding of exoplanet diversity. 
The NASA HWO \citep[][]{2021pdaa.book.....N} will pursue similar goals with a different technology and spectral coverage. Complementary observations from both facilities will yield higher scientific return than each would offer individually \citep{2024A&A...689A.245A}.

Both missions are developing stellar target catalogs to estimate exoplanet detection yields. 
For HWO, the preliminary catalogs are the System Properties \& Observational Reconnaissance for Exoplanet Studies \citep[SPORES,][]{2024ApJS..272...30H} and the HWO Preliminary Input Catalog \citep[HPIC,][]{2024AJ....167..139T}. 
Early versions of potential stellar target catalogs for LIFE were presented by \citet{2022A&A...664A..21Q} and \citet{RWIIIMenti}. 
As target catalogs depend, amongst other things, on (unfinalized) mission architectures, the LIFE team decided to create a large database of potential targets to facilitate the creation of different target lists satisfying various choices of target selection criteria. 
This note reports on the progress of this database, which is accessible online\footnote{\url{https://dc.zah.uni-heidelberg.de/voidoi/q/lp/custom/10.21938/ke_e6lzO_jjX_vzvVIcwZA}}.

The LIFE team has defined three sets of potential targets. All of these target sets are derived from the full database which contains as much data as possible on $\sim$$10^4$ systems of stars within 30\,pc of the Sun (no single brown dwarfs or white dwarfs, no magnitude cut). The first set is the input target catalog, dubbed LIFE-StarCat, which consists of $\sim$$10^3$ main sequence single stars and wide binaries which could have stable planet orbits in their habitable zone. LIFE-StarCat contains the input needed by {\tt LIFEsim} \citep{2022A&A...664A..22D}, a tool for simulating LIFE observations. The second set, the final LIFE target list, contains the $\sim$$10^2$ stars that, according to the {\tt LIFEsim} output, promise the highest detection yield fulfilling the scientific goals of the mission. The last set, with $\sim$$10^1$ objects, is the list of ``LIFE Golden Targets'', i.e., the best a-priori known nearby planetary systems.

\section{LIFE Target Database Features}

The primary use case of the database is the exploration of selection criteria for the creation of the input target catalog.
The use of LIFE-StarCat with {\tt LIFEsim}
allows for the prediction of exoplanet detection yields, specifically those for LIFE's primary science objective, which is to characterize 
dozens of 
Earth-size planets ($R_p \approx$ 0.5--1.5\,R$_\oplus$) in the empirical habitable zone \citep{2022ExA....54.1197Q}.

The yield estimates allow us to explore trade-offs in the mission design (e.g., changes in collector size, instrument throughput, wavelength coverage) and quantify their scientific impact (e.g., number of detectable exoplanets). This requires efficient interactions between LIFE's science and instrument teams. An early outcome of this interaction was the distance cut at 30\,pc in the full database, which includes enough stars to fulfill LIFE's primary science objectives. 

\begin{figure}[ht!]
\plotone{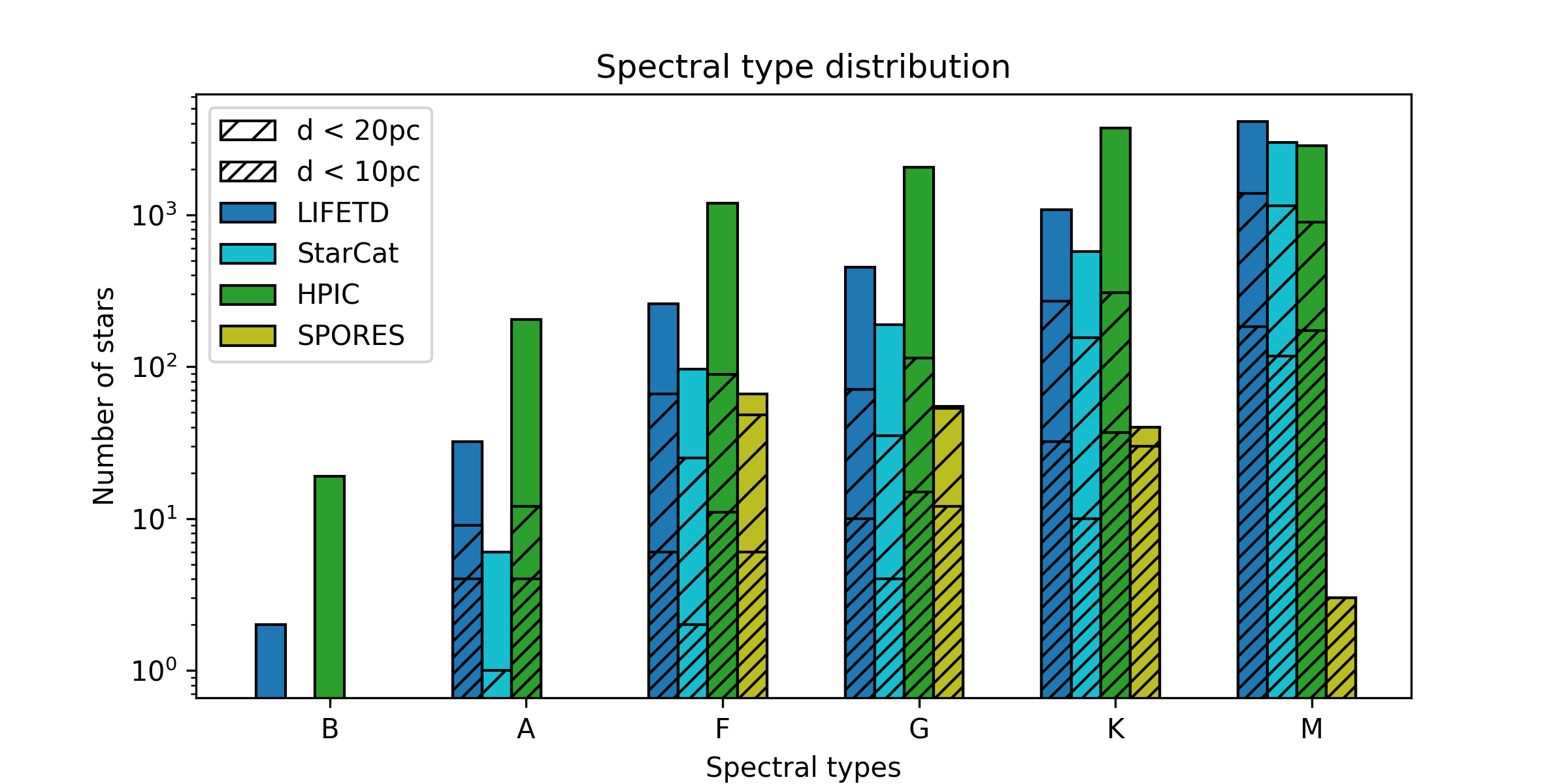}
\caption{
Spectral distribution of stars in the LIFE target database (dark blue), LIFE-StarCat (light blue),
HPIC (green), and SPORES (olive).
\label{fig:general}}
\end{figure}

Apart from the main stellar parameters needed for the yield estimates (e.g., equatorial coordinates, spectral type, mass), we compiled information on system multiplicity, debris disks, and planet (non-)detections.
The star's environment is essential to quantify the impact of background noise from exozodiacal dust or from close companions and, therefore, to determine its suitability as a potential LIFE target. 
Figure~\ref{fig:general} shows the current number of stars in LIFE-StarCat as a function of spectral type in comparison to the LIFE target database, SPORES, and HPIC.\\

The Virtual Observatory (VO)\footnote{\url{www.ivoa.net}} develops standards to easily interact with astronomical data. 
As our database is already VO-compatible, users can interact with already established VO tools (e.g., VizieR, Aladin, TOPCAT). 
Access to our database is available through the Table Access Protocol (TAP)\footnote{\url{http://dc.zah.uni-heidelberg.de/tap}} using PyVO or TOPCAT. The data can be extracted using the Astronomical Data Query Language (ADQL)\footnote{See examples at \url{https://dc.zah.uni-heidelberg.de/life/q/ex/examples}}. 
In addition, we provide the extraction and post-processing of the data leading to LIFE-StarCat as a Jupyter notebook tutorial\footnote{\url{https://life-td.readthedocs.io}}, from which one can generate user-defined catalogs.\\

We publish our LIFE target database through GAVO\footnote{\url{http://dc.zah.uni-heidelberg.de/browse/life/q}}, a reliable data center.
For each of the parameters in the database we reference the original provider (where available, we also store its original publication). We prioritize the data provider to minimize the workload of updating the database and a small number of providers supply most of the data in the LIFE target database. 
We use VO-compatible services as input providers when possible, such as SIMBAD \citep{2000A&AS..143....9W}, 
\textit{Gaia} \citep{2023A&A...674A...1G}, 
Exo-MerCat \citep{2020A&C....3100370A}, 
or the Washington Double Star Catalog accessed via the VizieR data center. 
When this option is unavailable, we use other non-VO-compliant databases, like the ``star database + sed models'' (sdb)\footnote{\url{http://drgmk.com/sdb/}}, maintained by G.~Kennedy, or derive the data ourselves.
Only after these steps we consider adding individual measurements from the literature. 
The benefit of this order is that ingesting new data from VO-compliant servers is automated and leads to a coherent dataset, while non-VO-compliant ones may need some adjustments in the ingestion pipeline. \\

Even considering multiple providers, there are promising targets for which we still lack essential data. One use case is to identify those gaps in knowledge so that forthcoming observation proposals can address them.
Our overarching goal for the database is to provide lasting context for the analysis of data from the LIFE mission. This means that the database shall remain useful and up-to-date over the next few decades. 
The LIFE target database structure was designed to be flexible and capable of adapting to changes in use cases that may arise in the future. 

\begin{acknowledgments}
We acknowledge the Workaut / ETH Z\"urich cooperation.
\end{acknowledgments}

\bibliography{life_td}{}
\bibliographystyle{aasjournal}

\end{document}